\documentclass[10pt]{iopart}
\usepackage[square,numbers]{natbib}

\usepackage{amsmath}
\usepackage{amssymb}
\usepackage{xcolor}
\usepackage{bm}
\usepackage{graphicx}
\usepackage{subcaption}
\usepackage{mathtools}
\usepackage{datetime}

\renewcommand\vec[1]{\mathbf{#1}}

\newdateformat{monthyeardate}{%
  \monthname[\THEMONTH] \THEYEAR}

\begin{document}

\letter{Stellarator coil optimization supporting multiple magnetic configurations}

\author{Brandon F Lee$^{1,2,3}$, Elizabeth J Paul$^{1,4}$, Georg Stadler$^5$, and Matt Landreman$^6$}

\address{$^1$ Princeton Plasma Physics Laboratory, 100 Stellarator Road, Princeton, NJ 08536}
\address{$^2$ Department of Biomedical, Biological, and Chemical Engineering, University of Missouri, 416 South 6\textsuperscript{th} Street, Columbia, MO 65211}
\address{$^3$ Department of Physics and Astronomy, University of Missouri, 701 South College Avenue, Columbia, MO 65211}
\address{$^4$ Department of Astrophysical Sciences, Princeton University, 4 Ivy Lane, Princeton, NJ 08544}
\address{$^5$ Courant Institute of Mathematical Sciences, New York University, 251 Mercer Street, New York, NY 10012}
\address{$^6$ Institute for Research in Electronics and Applied Physics, University of Maryland, 8223 Paint Branch Drive, College Park, MD 20742}
\eads{\mailto{brandonlee@mail.missouri.edu}, \mailto{epaul@princeton.edu}, \mailto{stadler@cims.nyu.edu}, \mailto{mattland@umd.edu}}
\vspace{10pt}
\begin{indented}
\item[]\monthyeardate\today
\end{indented}

\begin{abstract}
    We present a technique that can be used to design stellarators with a high degree of experimental flexibility. 
    For our purposes, flexibility is defined by the range of values the rotational transform can take on the magnetic axis of the vacuum field while maintaining satisfactory quasisymmetry.
    We show that accounting for configuration flexibility during the modular coil design improves flexibility beyond that attained by previous methods.
    Careful placement of planar control coils and the incorporation of an integrability objective enhance the quasisymmetry and nested flux surface volume of each configuration.
    We show that it is possible to achieve flexibility, quasisymmetry, and nested flux surface volume to reasonable degrees with a relatively simple coil set through an NCSX-like example.
    This example coil design is optimized to achieve three rotational transform targets and nested flux surface volumes in each magnetic configuration larger than the NCSX design plasma volume.
    Our work suggests that there is a tradeoff between flexibility, quasisymmetry, and volume of nested flux surfaces.
\end{abstract}

%
\vspace{2pc}

\submitto{\NF}
%
%
\ioptwocol

\section{Introduction}
Effective stellarator coil optimization is of great interest to the nuclear fusion community. Traditionally, a two-stage optimization process is used: a magnetic field with desirable physical properties is designed, then coils subject to reasonable engineering constraints are optimized to create the target field \cite{stellopt,coilopt}.
During the first stage, a ``fixed-boundary'' approach is used, which means that the shape of the last closed flux surface acts as a computational boundary during the magnetic equilibrium calculations. 
It is challenging to use the two-stage approach to design stellarators that can achieve more than one magnetic configuration with the same modular coil set because the stages are decoupled from one another. Many devices use control coils to explore several configurations, but these coils are not accounted for during the modular coil design. Designing a stellarator for flexibility from the outset would presumably make the device cheaper, and therefore more experimentally useful, than multiple devices with limited flexibility.

``Free-boundary'' equilibrium calculations could, in principle, be used to circumvent the difficulties introduced by the stage decoupling inherent in the two-stage approach. In the free-boundary approach, the coil geometries and currents are first specified, then the shape of the last closed flux surface is calculated self-consistently.
One can theoretically optimize coils to generate a desirable magnetic field by performing a free-boundary equilibrium calculation within each step of the coil optimization. In practice, however, such optimizations typically require substantial computing time and often struggle to converge.
Our method, outlined in section \ref{sec:formulation}, is somewhat similar to the free-boundary approach but uses a near-axis expansion to reduce computational complexity.

To achieve experimental flexibility, several stellarators have included planar coils to modify their magnetic configurations. While planar coils cannot provide intricate shaping like modular coils, they are relatively simple to manufacture and can make small modifications to the magnetic field. For instance, the design of the National Compact Stellarator Experiment (NCSX) included 18 identical, equally-spaced toroidal field coils to keep the mean toroidal field constant and free the modular coils for 3D shaping, as well as 4 poloidal field coils to adjust the plasma position and shape \cite{ncsxFlexibility,NCSXconfigDesign}. 
The currents in each modular and poloidal field coil could be modified independently, while all the toroidal field coils were intended to have one power supply.
Modifying these currents could change the value of the rotational transform, $\iota$, on the magnetic axis from approximately $0.175$ to $0.5$ while maintaining constant shear.

The original design of the Helically Symmetric Experiment (HSX) did not include planar coils. To degrade the quasisymmetry of the field and improve device flexibility, planar coils were fastened to the plates that already supported the modular coils \cite{HSXcoilDesign}. Changing the direction of the currents in the planar coils independently allowed for the rotational transform on the axis to be changed by about $10\%$ and significantly degraded the quasisymmetry of the device \cite{HSXcontrolCoilPerformance}. 

As with HSX, the initial coil design for Wendelstein 7-X (W7-X) did not include planar coils \cite{w7xNewerCoilDesign}. However, due to the substantial experimental flexibility offered by the planar coils in the Wendelstein 7 Advanced Stellarator (W7-AS), which allowed for the rotational transform to take values between $0.25$ and $0.7$, W7-X was redesigned to include planar coils \cite{w7xOlderCoilDesign}. 
The W7-X standard configuration, corresponding to no current in the planar coils, has an edge rotational transform of $5/5$. Running current through the planar coils can change the edge rotational transform to $5/4$ or $5/6$ \cite{w7xNewerCoilDesign,w7xConfig1}. 
The planar coils can also be used to affect several other characteristics of the magnetic field and plasma \cite{w7xNewerCoilDesign,w7xConfig1,W7XcoilDesign,w7xConfig2}.

The modular coils of the Quasi-Poloidal Stellarator (QPS) were designed to achieve a single magnetic configuration. 
Independent power supplies were envisioned for each modular coil and three pairs of vertical field coils, while all twelve toroidal field coils would have used one power supply. The plasma current could also be manipulated with an ohmic transformer \cite{QPSflexibility}. Optimizations of the aforementioned currents resulted in on-axis rotational transform values of roughly 0.165 to 0.375 with less than a $15\%$ change in the ion energy confinement.

In this work we utilize the coil design approach proposed in \cite{PPO}, which couples the vacuum magnetic field and coil optimizations together by incorporating objectives from both stages into a single objective function so that they can be optimized simultaneously in one step.
This is different from the two-stage design approach discussed above, which requires sequential optimization of the magnetic field and coil objective functions.
The approach of \cite{PPO}, which is implemented in the PyPlasmaOpt software, employs a near-axis expansion \cite{QS1,QS2,QS3} to achieve quasisymmetry in the neighborhood of the magnetic axis. Using this expansion makes the calculations similar in spirit to free-boundary coil optimizations, but much less computationally intensive.
As previously mentioned, some experimental devices have used planar coils to change the rotational transform and intentionally degrade quasisymmetry, but achieving configuration flexibility and quasisymmetry simultaneously is challenging. 
We modify PyPlasmaOpt to take flexibility and planar coil placement into account during the initial optimization rather than first optimizing the stellarator for a standard state and adding planar coils after the fact. We also compute quadratic flux minimizing (QFM) surfaces \cite{qfm} throughout the optimization to improve the integrability of the field away from the magnetic axis. 
For the purposes of demonstration, we use approximate design parameters of NCSX as the starting point for our optimizations. 
We show that device flexibility can be improved by accounting for it in the coil design stage. 
An example device is presented that can achieve a range in $\iota$ on the axis, nested flux surface volumes in each configuration larger than the design plasma volume of NCSX, and quasisymmetry in each configuration of similar magnitude to the NCSX design.
Our quasisymmetry optimization approach relies on minimizing the difference between the total magnetic field and a target field on the magnetic axis, making it easy to use when only vacuum fields are considered. However, it would not be straightforward to extend this approach to finite-$\beta$ plasmas because the plasma current would need to be known throughout the device (as in free-boundary optimizations) to determine the total magnetic field.
Our general approach to optimizing for flexibility, on the other hand, could readily be extended beyond vacuum fields.

\section{Formulation}
\label{sec:formulation}
The basic idea of this work is that flexibility can be achieved by incorporating several objective functions, one for each of the desired magnetic configurations, into a single composite objective. We assume that flexibility is realized by changing the currents in each unique coil while holding the coil shapes constant. One can imagine that the composite objective for $N_{\mathrm{m}}$ desired magnetic configurations would take the general form
\begin{equation}
    f_{\mathrm{comp}} \coloneqq f_{\mathrm{indep}} + \sum_{j=1}^{N_{\mathrm{m}}} f_{\mathrm{dep},j},
    \label{eqn:generic}
\end{equation}
where $f_{\mathrm{indep}}$ is an objective function that is independent of the coil currents and $f_{\mathrm{dep,j}}$ are objective functions that are dependent on the coil currents. In this work, $f_{\mathrm{indep}}$ contains coil regularization terms and the $f_{\mathrm{dep,j}}$ contain terms related to the rotational transform, quasisymmetry, and nested flux surface volume of each magnetic configuration. The remainder of this section describes the details of our formulation.

PyPlasmaOpt calculates a quasiaxisymmetric field in the neighborhood of the magnetic axis and optimizes coils such that they create a vacuum field on the axis that coincides with the quasisymmetric field. Magnetic field contributions from the plasma current are not taken into account in this model, eliminating the need to solve the more expensive free-boundary equilibrium problem.
The coils are subject to a regularization objective determined from engineering considerations.

The coil regularization objective is 
\begin{multline}
    R(\vec{c}) \coloneqq \sum_{i=1}^{N_{\mathrm{c}}} \left[
    \frac{1}{2}  \left(\frac{L_{\mathrm{c}}^{(i)}(\vec{c})-L_{\mathrm{c}}^{(i,\mathrm{target})}}{L_{\mathrm{c}}^{(i,\mathrm{target})}}\right)^{2} \right.\\
    + \alpha \int \mathrm{max}\left(0,\kappa_{i}(\vec{c})-\frac{2\pi}{L_{\mathrm{c}}^{(i)}(\vec{c})}\right)^{4} d\ell_{i} \\
     + \left. \gamma \sum_{k=1}^{i-1} \int \int \mathrm{max}\left(0,d_{\mathrm{min}}-\|\vec{r}_{i}(\vec{c})-\vec{r}_{k}(\vec{c})\|\right)^{2} d\ell_{i} d\ell_{k} \right],
     \label{eqn:regTerms}
\end{multline}
where $N_c$ is the total number of unique coils, $L_{\mathrm{c}}^{(i)}$ is the length of coil $i$, $\kappa_i$ is the curvature of coil $i$, $d_{\mathrm{min}}$ is the minimum desired intercoil distance, $\vec{r}_i$ is the position vector along coil $i$, and $\alpha$ and $\gamma$ are weights. These regularization terms prevent the length and shape of the coils from becoming physically unrealistic. For each optimization, we set $\alpha=10^{-6}$, $\gamma=1000$, and $d_{\mathrm{min}}=0.2$ similar to \cite{PPO}.

Some objective function terms unique to magnetic configuration $j$ are
\begin{multline}
    f_{\mathrm{config,}j}(\vec{c},\vec{a}) \coloneqq  \frac{\int_{\mathrm{MA},j} \|\vec{B}_{\mathrm{coils},j}\left(\vec{c}\right)-\vec{B}_{\mathrm{QS},j}\left(\vec{a}\right)\|^{2} d\ell}{\int_{\mathrm{MA},j}\|\vec{B}_{\mathrm{coils},j}\left(\vec{c}\right)\|^{2} d\ell} \\
    + \frac{\int_{\mathrm{MA},j} \|\nabla\vec{B}_{\mathrm{coils},j}\left(\vec{c}\right)-\nabla\vec{B}_{\mathrm{QS},j}\left(\vec{a}\right)\|^{2} d\ell}{\int_{\mathrm{MA},j} \|\nabla\vec{B}_{\mathrm{coils},j}\left(\vec{c}\right)\|^{2} d\ell} \\
     + \frac{1}{2} \left(\frac{\iota_{j} \left(\vec{a}\right)-\iota^{\mathrm{target}}_{j}}{\iota^{\mathrm{target}}_{j}}\right)^{2} + \frac{1}{2} \left(\frac{L_{\mathrm{a},j}\left(\vec{a}\right)-L^{\mathrm{target}}_{\mathrm{a},j}}{L^{\mathrm{target}}_{\mathrm{a},j}}\right)^{2},
     \label{eqn:uniqueTerms}
\end{multline}
where $\vec{c}$ are the coil parameters, $\vec{a}$ are the magnetic axis parameters, $\vec{B}_{\mathrm{coils}}$ is the magnetic field produced by the coils, $\vec{B}_{\mathrm{QS}}$ is the target quasisymmetric field, $L_{\mathrm{a}}$ is the length of the magnetic axis, and $\| \cdot \|$ denotes the $L^2$-norm for vectors and the Frobenius norm for tensors.
The quasisymmetry model takes a specified magnetic axis and a parameter related to the elongation of the flux surfaces as its inputs and returns $\vec{B}_{\mathrm{QS}}$, $\nabla \vec{B}_{\mathrm{QS}}$, and $\iota$ by numerically solving a first-order nonlinear ordinary differential equation \cite{QS1,QS2,QS3}. This means that the magnetic axis used by the quasisymmetry model and the true magnetic axis from the coils do not necessarily coincide unless the error in the first two terms of the objective \eqref{eqn:uniqueTerms} is small. We also rely on small error in these terms for an accurate calculation of $\iota$. 
The third term in equation \eqref{eqn:uniqueTerms} enforces our target value of $\iota$ by manipulating the magnetic axis fed to the quasisymmetry model, and the fourth prevents the magnetic axis from becoming very large or small.  

As discussed in section~\ref{sec:notMagicSchemes}, we found that optimizing for multiple $\iota^{\mathrm{target}}$ values could create smaller-than-desired volumes of nested flux surfaces. In an attempt to mitigate this, we compute a QFM surface \cite{qfm} for each magnetic configuration throughout the optimization. A QFM surface with a unit normal vector $\hat{\vec{n}}_{j}$ is defined as the minimizer of the objective function 
\begin{equation}
    f_{\mathrm{QFM},j}(\vec{S},\vec{c}) \coloneqq \frac{\int_{\mathrm{S},j} \left(\vec{B}_{\mathrm{coils},j} \cdot \hat{\vec{n}}_{j}\right)^{2} d^{2}x}{\int_{\mathrm{S},j} B_{\mathrm{coils},j}^{2} d^{2}x}
    \label{eqn:qfmObj}
\end{equation}
with respect to the surface parameters, $\vec{S}$, subject to the constraint that the volume equals a user-specified constant.
The resulting QFM surface is an approximate magnetic surface, and the value of $f_{\mathrm{QFM},j}$  quantifies the deviation of the field from integrability. Each QFM surface is recomputed at every iteration of the optimization, and the value of equation~\eqref{eqn:qfmObj} is then penalized as shown in equation \eqref{eqn:PPOobjective}. 

We enable the code to simultaneously optimize $N_{\mathrm{m}}$ magnetic configurations that share the same coil set, each with their own $\iota^{\mathrm{target}}$, by implementing the composite objective
\begin{multline}
    f_{\mathrm{comp}}(\vec{c},\vec{a}) \coloneqq \\
    R(\vec{c}) + \sum_{j=1}^{N_{\mathrm{m}}} \left(f_{\mathrm{config,}j}(\vec{c},\vec{a})+\eta \min_{\vec{S}} f_{\mathrm{QFM},j}(\vec{S},\vec{c})\right),
    \label{eqn:PPOobjective}
\end{multline}
where $\eta$ is a weight.

The variation of each $f_{\mathrm{QFM},j}$ with respect to $\vec{c}$ is
\begin{multline}
    \delta f_{\mathrm{QFM},j} (\delta\vec{c};\vec{S(\vec{B})},\vec{B}) = \\
    \delta f_{\mathrm{QFM},j} (\delta \vec{B};\vec{S},\vec{B}) + \delta f_{\mathrm{QFM},j} (\delta \vec{S}(\delta \vec{B});\vec{S(\vec{B})},\vec{B}),
    \label{eqn:f_qfmDer}
\end{multline}
where $\delta \vec{B} = \delta \vec{B} (\delta \vec{c})$.
The first term in equation~\eqref{eqn:f_qfmDer} corresponds to the explicit dependence of each $f_{\mathrm{QFM},j}$ on the magnetic field, and the second term corresponds to the implicit dependence of each $f_{\mathrm{QFM},j}$ on the magnetic field through the QFM surface shape.
Because each $f_{\mathrm{QFM},j}$ is evaluated at a stationary point with respect to $\vec{S}$, only the first term of equation~\eqref{eqn:f_qfmDer} contributes.
The derivatives of the other terms in the composite objective function \eqref{eqn:PPOobjective} are computed analytically or using adjoint methods as described in \cite{PPO}.
The value of the objective is minimized by the L-BFGS-B algorithm \cite{lbfgs1,lbfgs2} implemented in SciPy \cite{scipy}.

To quantify the quasiaxisymmetry (QA) of an optimized configuration during postprocessing, we define the QA error to be
\begin{equation}
    \epsilon_{\mathrm{QA}} \coloneqq \sqrt{\frac{\sum_{m,n\neq 0} B_{m,n}^{2}}{\sum_{m,n} B_{m,n}^{2}}},
    \label{eqn:QAmetric}
\end{equation}
where $B_{m,n}$ refer to the Fourier harmonics of the magnetic field in Boozer coordinates on a given flux surface. These harmonics are obtained from BOOZ{\_}XFORM \cite{BOOZXFORM} using the vacuum field calculated from free-boundary VMEC \cite{VMEC}.

\section{Optimization for Flexible QA Configuration}
\label{sec:magic}
\begin{figure*}
    \centering
    \begin{subfigure}[b]{0.25\textwidth}
    \includegraphics[width=\linewidth]{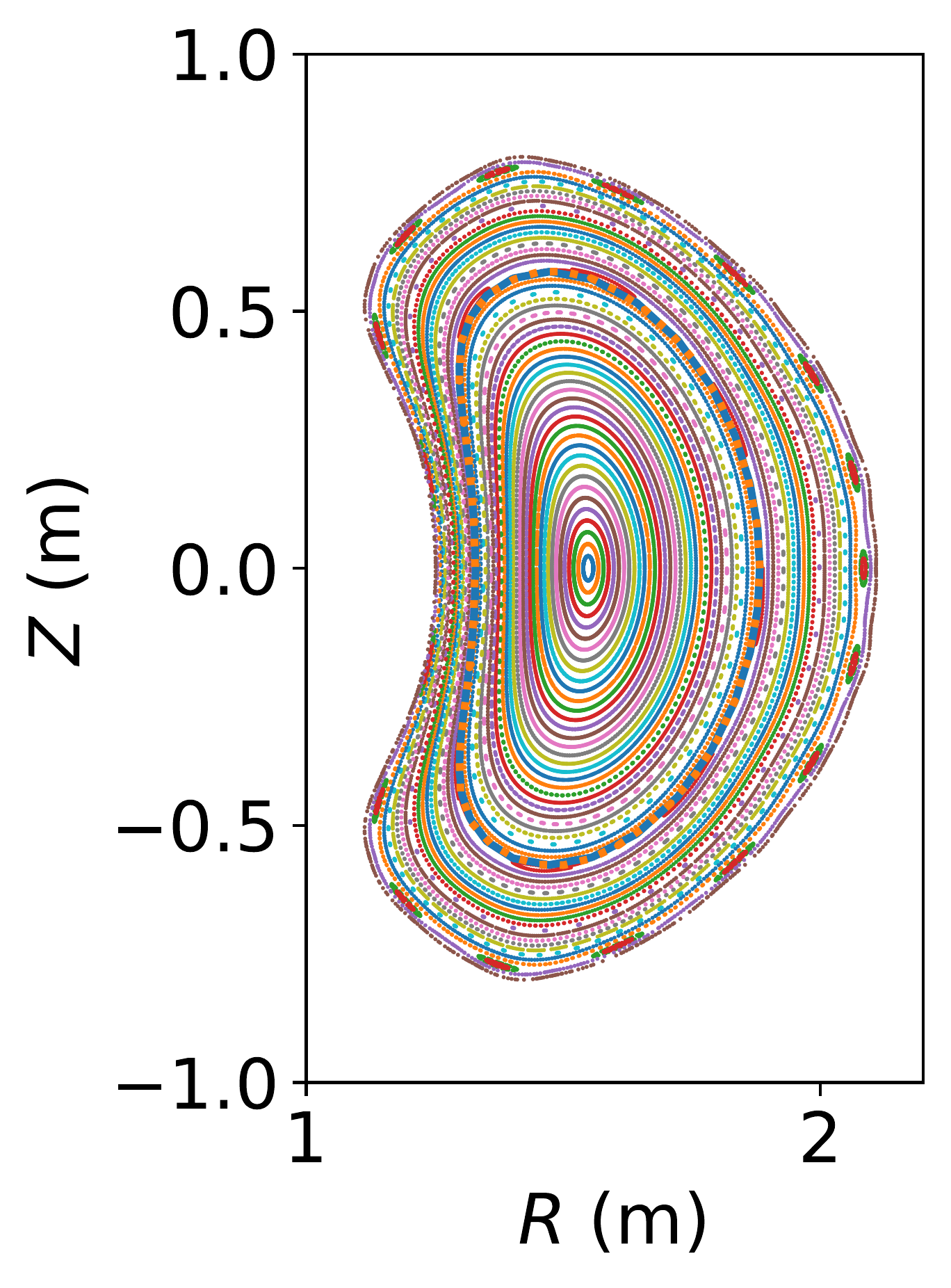}
    \caption{}
    \label{fig:magicPoincare1}
    \end{subfigure}
    \begin{subfigure}[b]{0.25\textwidth}
    \includegraphics[width=\linewidth]{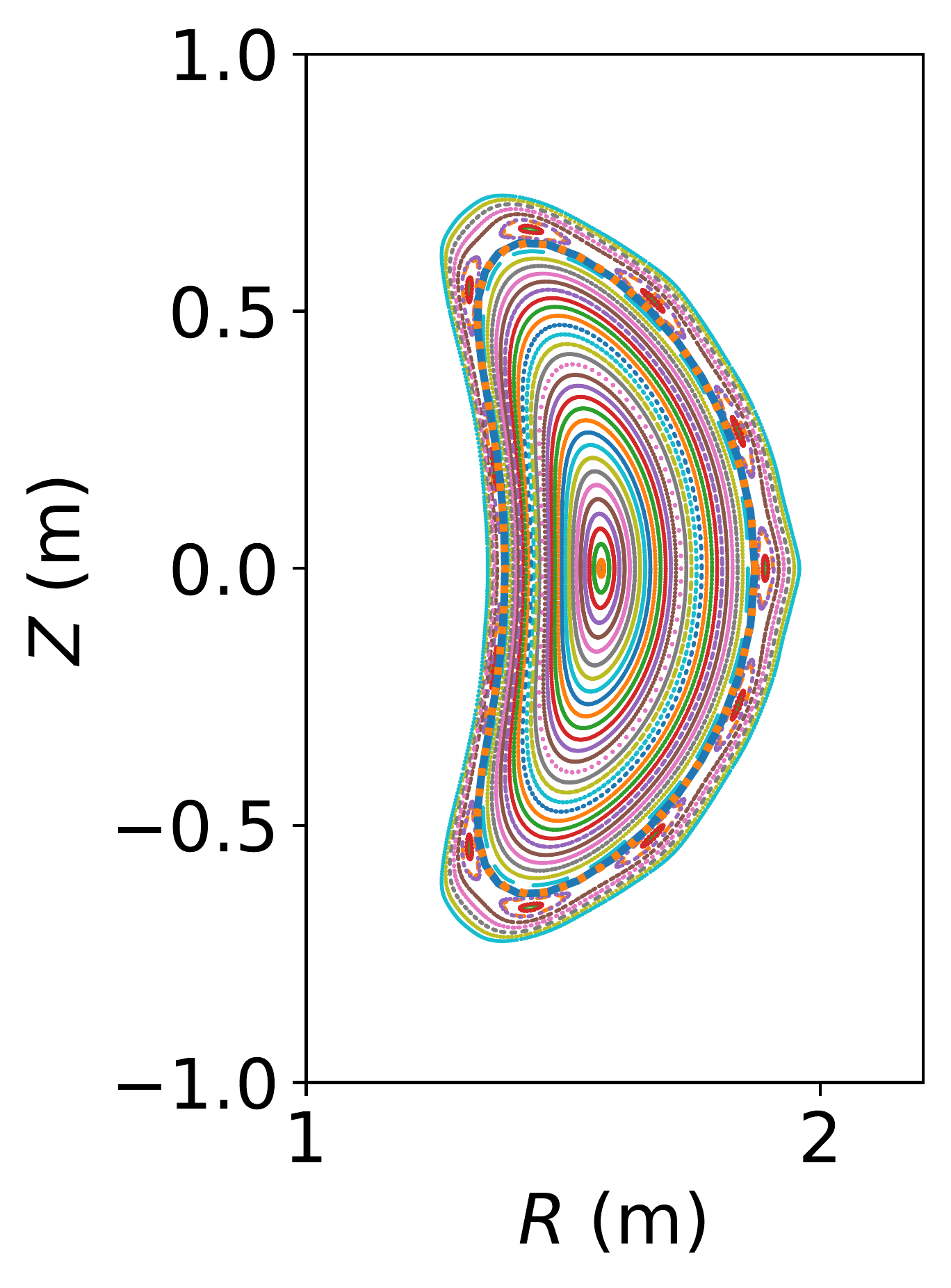}
    \caption{}
    \label{fig:magicPoincare2}
    \end{subfigure}
    \begin{subfigure}[b]{0.25\textwidth}
    \includegraphics[width=\linewidth]{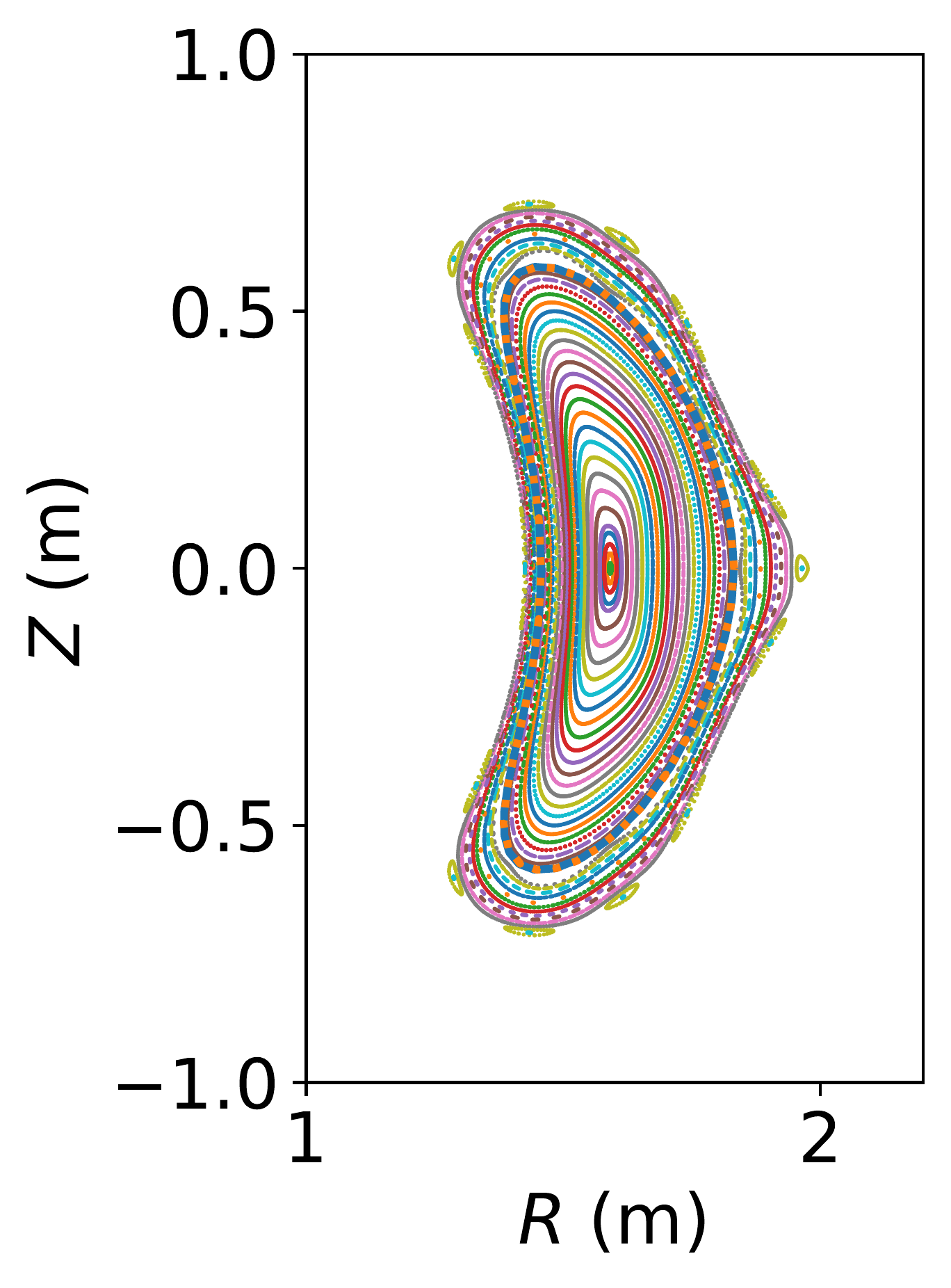}
    \caption{}
    \label{fig:magicPoincare3}
    \end{subfigure}
    \begin{subfigure}[b]{0.33\textwidth}
    \includegraphics[width=\linewidth]{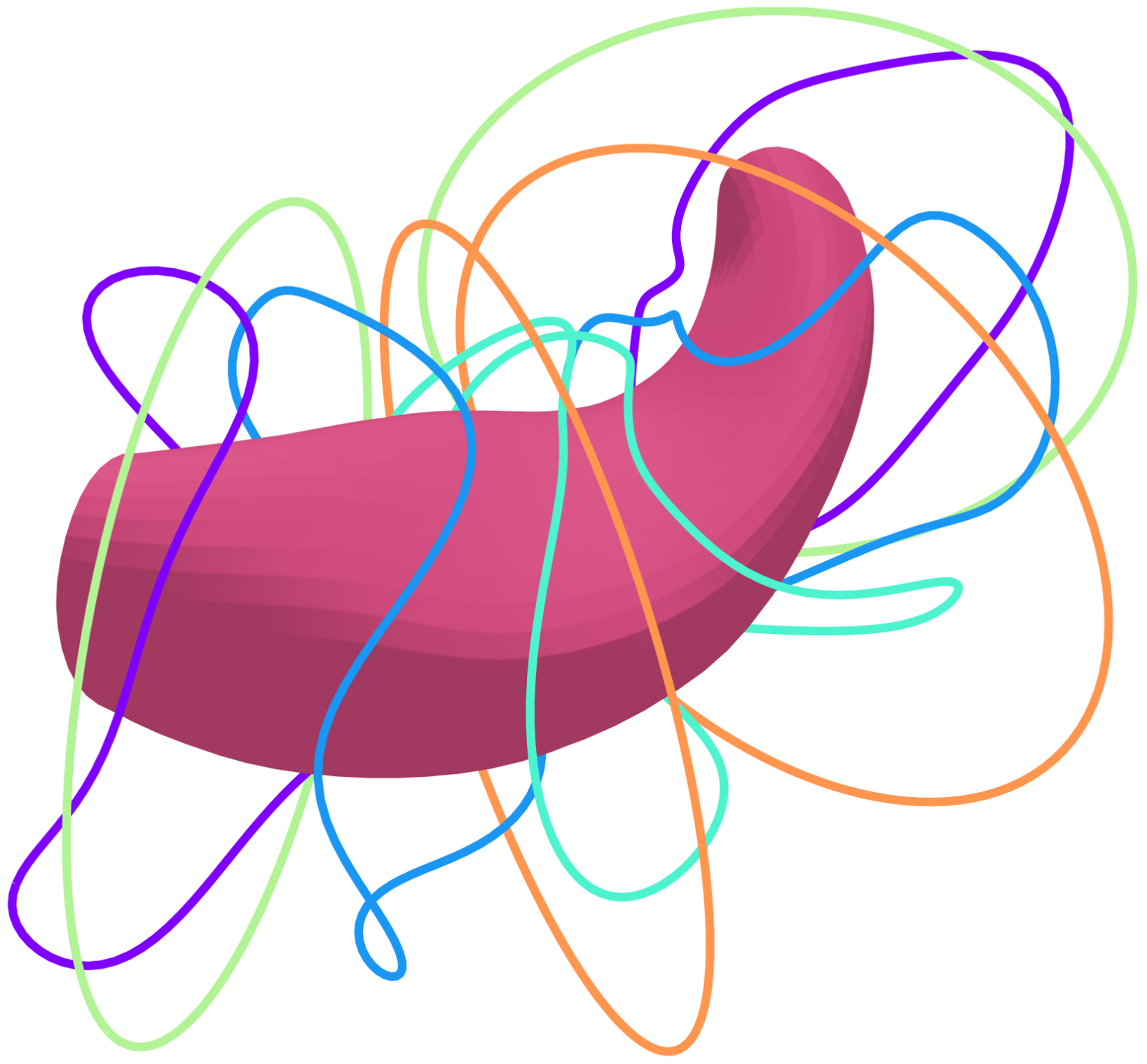}
    \caption{}
    \label{fig:magicAllCoils}
    \end{subfigure}
    \begin{subfigure}[b]{0.33\textwidth}
    \includegraphics[width=\linewidth]{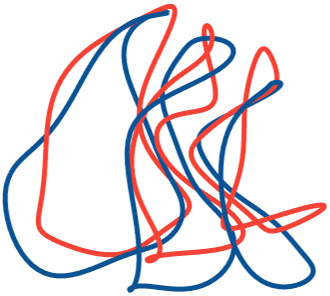}
    \caption{}
    \label{fig:magicUniqueCoils}
    \end{subfigure} \\
    \begin{subfigure}[b]{0.35\textwidth}
    \includegraphics[width=\linewidth]{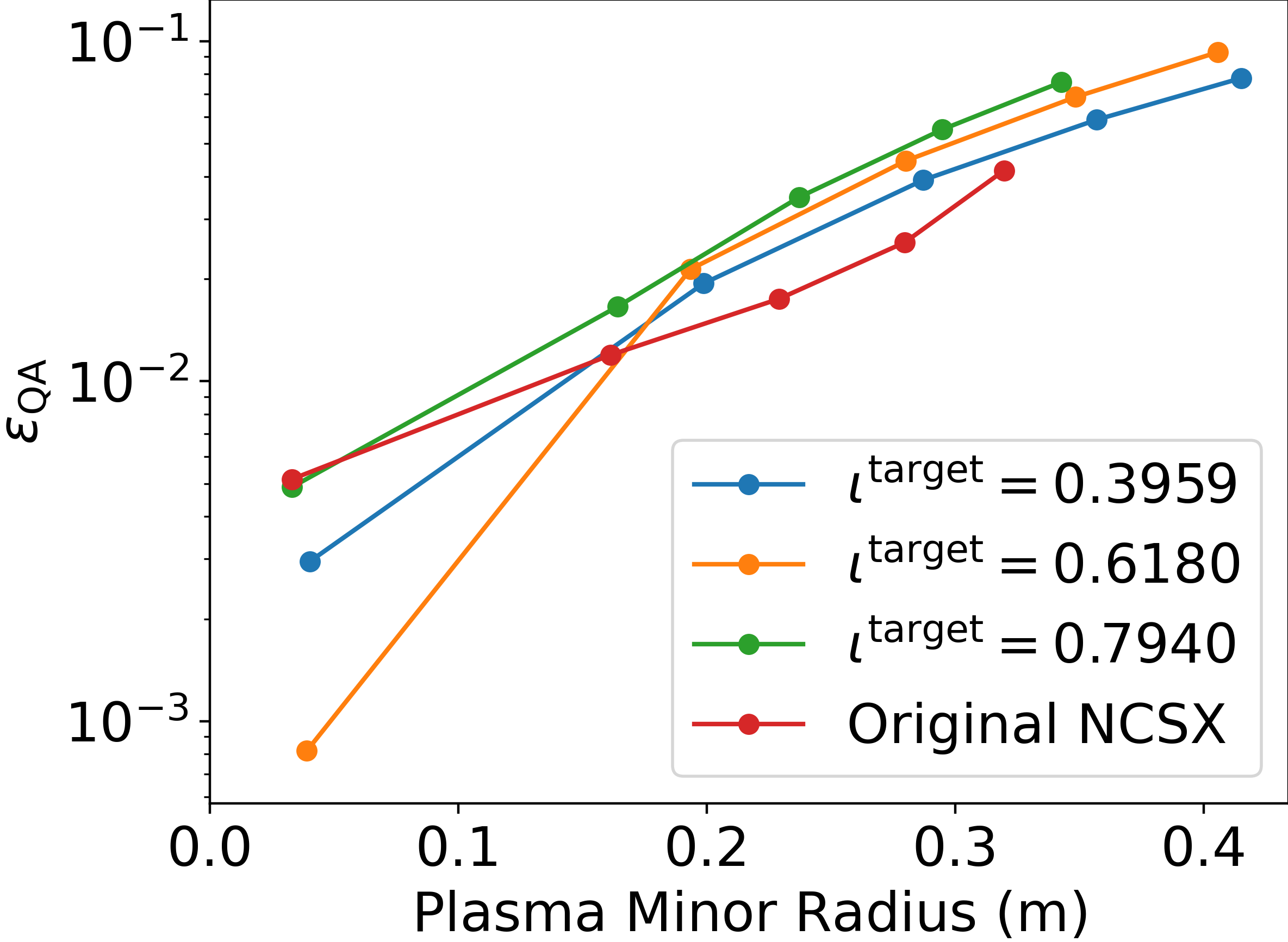}
    \caption{}
    \label{fig:magicQA}
    \end{subfigure}
    \begin{subfigure}[b]{0.35\textwidth}
    \includegraphics[width=\linewidth]{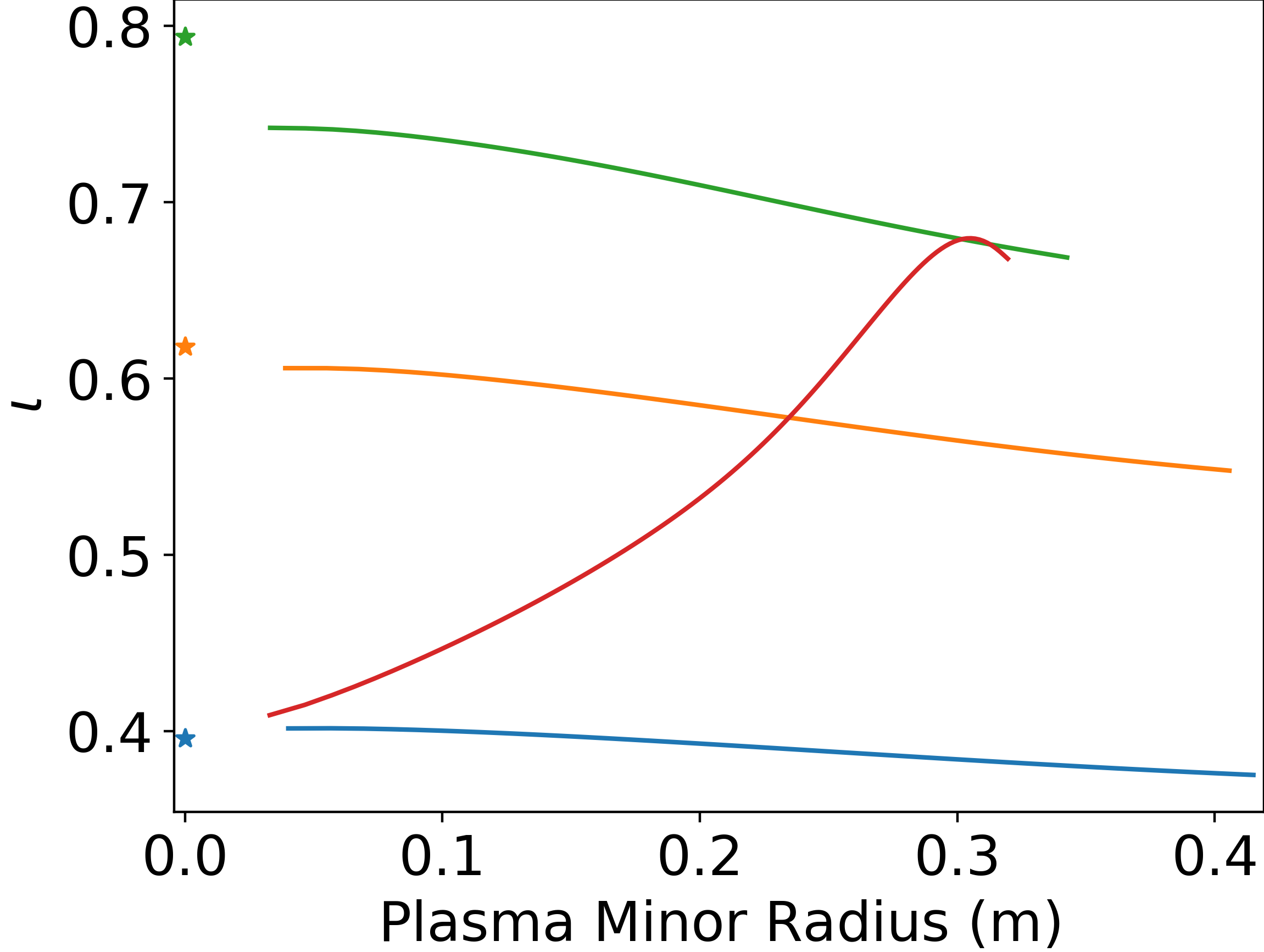}
    \caption{}
    \label{fig:magicIota}
    \end{subfigure}
    \caption{(\subref{fig:magicPoincare1}) (\subref{fig:magicPoincare2}), and (\subref{fig:magicPoincare3}) show Poincar\'e plots for the $\iota^{\mathrm{target}} = 0.3959$, $\iota^{\mathrm{target}} = 0.6180$, and $\iota^{\mathrm{target}} = 0.7940$ configurations, respectively, of the flexible device discussed in section~\ref{sec:magic}. The (overlapping) solid blue and dotted orange curves indicate the QFM and free-boundary VMEC surfaces, respectively, used during postprocessing. Twelve poloidal and toroidal modes were used for the VMEC calculations. The volumes of the VMEC surfaces shown are $5.166$~$\mathrm{m^3}$ (\subref{fig:magicPoincare1}), $4.954$~$\mathrm{m^3}$ (\subref{fig:magicPoincare2}), and $3.534$~$\mathrm{m^3}$ (\subref{fig:magicPoincare3}).
    Note that the flux surfaces become elongated as $\iota$ increases, which is consistent with previous work \cite{QAlim1,QAlim3}. (\subref{fig:magicAllCoils}) shows the modular and planar coils of the flexible device over one field period and the VMEC boundary associated with (\subref{fig:magicPoincare1}). (\subref{fig:magicUniqueCoils}) shows a comparison between the modular coils of the flexible device (blue) and those of the original NCSX design (red). (\subref{fig:magicQA}) shows $\epsilon_{\mathrm{QA}}$ for each configuration of the flexible device and the original NCSX design \cite{ncsxFlexibility}. (\subref{fig:magicIota}) shows the rotational transform for each configuration of the flexible device and the original NCSX design calculated using free-boundary VMEC. Stars indicate the (on-axis) optimization targets. The effective minor radius in (\subref{fig:magicQA}) and (\subref{fig:magicIota}) is computed with the same definition used in VMEC, given in \cite{QS3}.}
    \label{fig:magic}
\end{figure*}

Several approaches were explored to develop an optimization formulation that allows for reasonable achievement of flexibility, quasisymmetry, and a large volume of nested flux surfaces. Some of these approaches are discussed in sections \ref{sec:frozen} and \ref{sec:notMagicSchemes}. For the successful formulation, we optimized for three values of $\iota^{\mathrm{target}}$: $0.3959$, $0.6180$, and $0.7940$. The first is the on-axis $\iota$ of the original NCSX design \cite{NCSXconfigDesign}, while the second and third are substantially larger noble irrationals. It is known that QA becomes progressively more difficult to achieve as $\iota$ grows \cite{QAlim1, QAlim2, QAlim3}, so choosing two large $\iota^{\mathrm{target}}$ values serves as a challenging test for our methods, especially since we only consider the vacuum field. We initialize the optimization using an approximation of the NCSX modular coils \cite{NCSXconfigDesign,NCSXcoilEngr} with six Fourier modes. We add two planar coils per half field period as well. 
The parameters of each planar coil included in the optimization space were its center, orientation angles, radius, and current. The planar coils were initialized as circles with radii of 1.3~m and placed equidistant around the device.
QFM surfaces with volume $2.959{\ }\mathrm{m^3}$ (the design NCSX plasma volume), $\eta=50$, three poloidal modes, and three toroidal modes were added to each magnetic configuration during the optimization. The number of poloidal and toroidal modes was raised to six or eight during postprocessing to obtain smoother surfaces, and the surface volumes were adjusted to achieve agreement with the Poincar\'e plots shown in figure~\ref{fig:magic}. These surfaces served as the initial guesses for the free-boundary VMEC calculations.
The coils produced by the optimization are shown in figure~\ref{fig:magicAllCoils}, and the currents can be found in table~\ref{tab:currents}.

\begin{table}[]
    \centering
    \caption{Optimized modular and planar coil currents (kA) needed to produce the results of section \ref{sec:magic}. Note that the magnetic field is normalized to have an average magnitude of 1~T on the axis.}
    \begin{tabular}{c|c|c|c|}
        & Small $\iota$ & Medium $\iota$ & Large $\iota$ \\
         \hline
        Mod.{\ }1 & 360.5 & 443.5 & 491.3 \\
        Mod.{\ }2 & 355.4 & 415.1 & 440.5 \\
        Mod.{\ }3 & 274.0 & 325.1 & 349.9 \\
        Plan.{\ }1 & -136.6 & -33.1 & 41.7 \\
        Plan.{\ }2 & -124.4 & -33.5 & -9.3 \\
    \end{tabular}
    \label{tab:currents}
\end{table}

\begin{table}
    \centering
    \caption{Comparison of the maximum curvature ($\mathrm{m^{-1}}$), minimum intercoil distance (m), and maximum length (m) for the modular coils in the original NCSX design and the example design presented in section \ref{sec:magic}.}
    \begin{tabular}{c|c|c|}
        & Orig.{\ }NCSX & Ex.{\ }Design \\
         \hline
        Max. $\kappa$ & 10.6 & 3.81 \\
        Min. Intercoil Dist. & 0.152 & 0.198 \\
        Max. Coil Len. & 7.29 & 8.17 \\
    \end{tabular}
    \label{tab:coilStatsComparison}
\end{table}

\begin{table*}
    \centering
    \caption{Currents (in kA-turns) in the modular, toroidal field, and poloidal field coils from the NCSX flexibility study of Pomphrey et al.~\cite{ncsxFlexibility} for comparison with the currents of the flexible device outlined in section~\ref{sec:magic}, which are shown in table~\ref{tab:currents}. Note that PF1 and PF2 were excluded from the flexibility calculations as they were primarily dedicated to ohmic heating and that the average toroidal field strength was $1.7$~T at $R=1.4~\mathrm{m}$.}
    \begin{tabular}{c|c|c|c|c|c|c|c|c|}
        & Mod.{\ }1 & Mod.{\ }2 & Mod.{\ }3 & TF & PF3 & PF4 & PF5 & PF6 \\
        \hline
        Target $\iota^{\mathrm{axis}}=0.50$ & 759.2 & 714.6 & 626.1 & -39.2 & 1524.2 & 159.0 & 208.2 & -1.3 \\
        $\iota^{\mathrm{axis}}=0.40$ (Ref. Config.) & 694.2 & 654.6 & 551.1 & 27.8 & 1524.2 & 1180.0 & 95.2 & -2.3 \\
        Target $\iota^{\mathrm{axis}}=0.30$ & 616.2 & 593.6 & 485.1 & 100.8 & 1524.2 & 787.0 & 62.2 & -2.3 \\
        Target $\iota^{\mathrm{axis}}=0.20$ & 513.2 & 497.6 & 390.1 & 194.8 & -159.8 & 233.0 & 55.2 & 4.7 \\
    \end{tabular}
    \label{tab:ncsxCurrents}
\end{table*}

Our results are best understood when compared to the NCSX flexibility study carried out by Pomphrey et al.~\cite{ncsxFlexibility}.
They performed several optimizations of the coil currents with fixed coil shapes and shear, seeking a different value of  $\iota$ during each optimization. They considered non-zero plasma $\beta$ and current. 
From figure~\ref{fig:magicUniqueCoils} and table~\ref{tab:coilStatsComparison}, it is clear that the coils in our example design are simpler according to our regularization criteria than those of the NCSX design. This suggests that they would be easier to construct.
Comparing the currents obtained by our optimization with those from the NCSX flexibility study, shown in table~\ref{tab:ncsxCurrents}, suggests that our currents are reasonable.
The Poincar\'e plots in figure~\ref{fig:magic} show that the volume of nested flux surfaces shrinks somewhat for configurations with larger $\iota$, similar to the previous study.
Even so, the nested flux surface volume for each configuration presented in this section is greater than the design NCSX plasma volume.
In the previous study, the minimum on-axis $\epsilon_{\mathrm{QA}}$ is $0.004873$ when $\iota^{\mathrm{axis}} = 0.49$. As shown in figure~\ref{fig:magicQA}, the maximum on-axis $\epsilon_{\mathrm{QA}}$ for our flexible device is $0.004877$ in the $\iota^{\mathrm{target}} = 0.7940$ configuration.
However, it should be noted that the original NCSX design had superior $\epsilon_{\mathrm{QA}}$ far from the axis, likely because this was not accounted for in our objective function.
Similarly, other desirable traits not included in our objective function may be absent from the flexible device. For instance, all three of its magnetic configurations have magnetic hills whereas the NCSX design has a magnetic well.
The trends in $\epsilon_{\mathrm{eff}}^{3/2}$ follow those in figure~\ref{fig:magicQA}. A plot comparing $\epsilon_{\mathrm{eff}}^{3/2}$ for the configurations described in this section and the original NCSX design, calculated using NEO \cite{NEO}, can be found in the supplementary material.
Figure~\ref{fig:magicIota} shows that we achieve on-axis rotational transforms in roughly the range $0.4$ to $0.74$, which is larger than the range of $0.175$ to $0.5$ in the previous study and more challenging to achieve in a QA configuration due to the large values of $\iota$. 
Taken together, the optimization presented in this work resulted in an example coil design that achieves substantial flexibility, quasisymmetry near the axis comparable to the original NCSX design, and a large volume of nested flux surfaces in all three targeted magnetic configurations. 
This was achieved with 12 carefully-placed planar toroidal field coils, whereas the original NCSX design utilized 18 equally-spaced toroidal field coils and 4 poloidal field coils.

\section{Sequential Optimization}
\label{sec:frozen}
\begin{figure*}
    \centering
    \begin{subfigure}[b]{0.22\textwidth}
    \includegraphics[width=\linewidth]{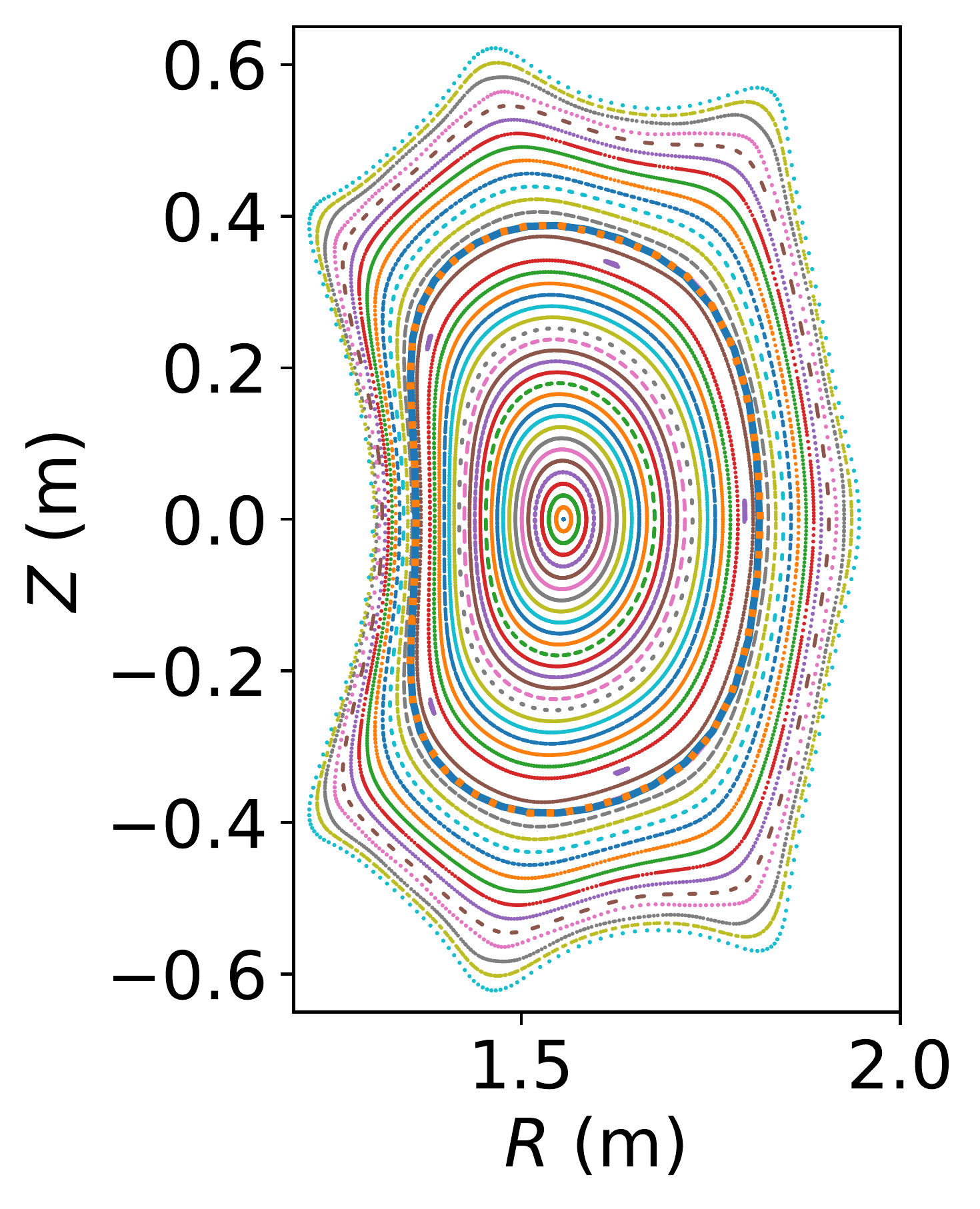}
    \caption{}
    \label{fig:frozenPoincare1}
    \end{subfigure}
    \begin{subfigure}[b]{0.22\textwidth}
    \includegraphics[width=\linewidth]{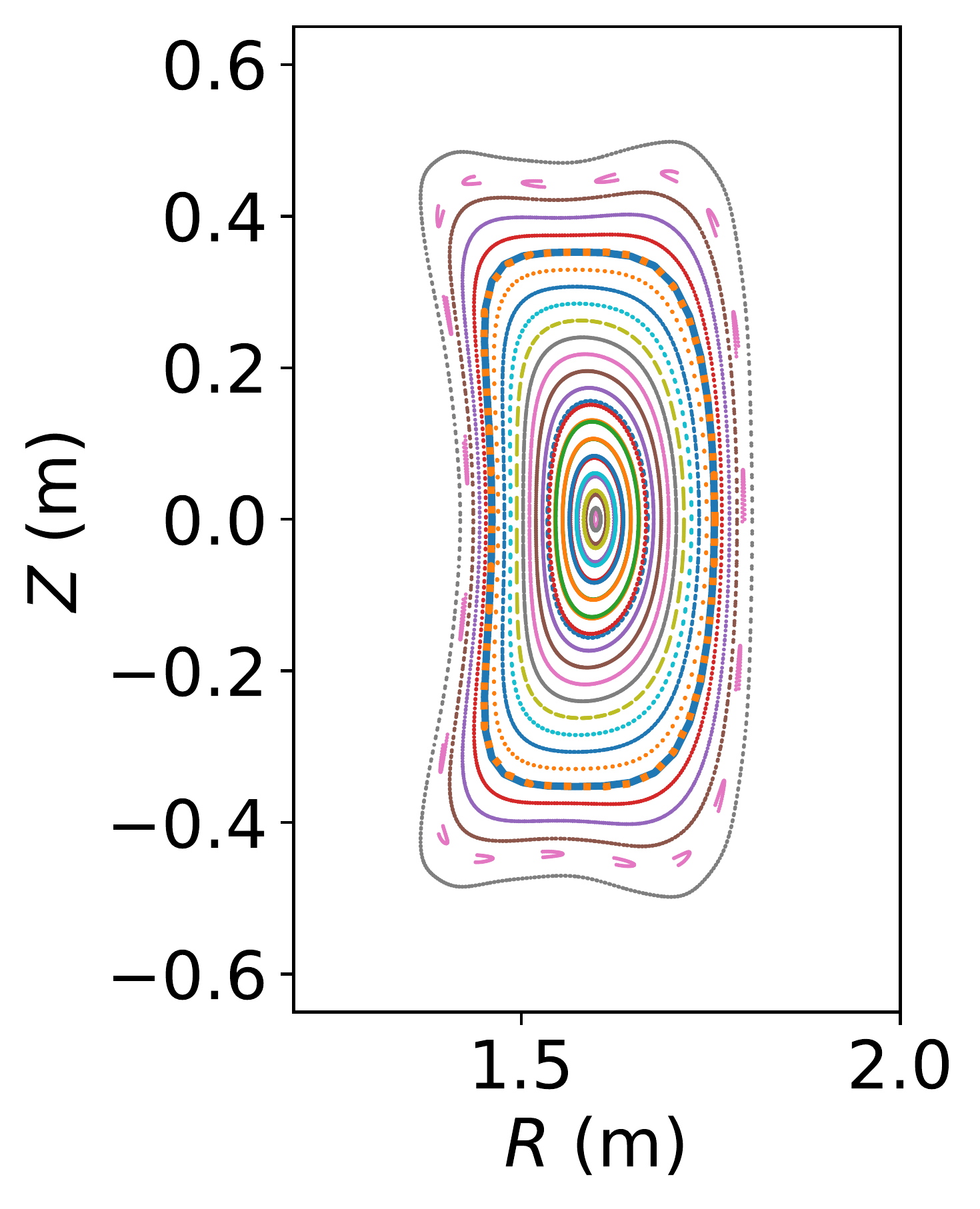}
    \caption{}
    \label{fig:frozenPoincare2}
    \end{subfigure} \\
    \begin{subfigure}[b]{0.25\textwidth}
    \includegraphics[width=\linewidth]{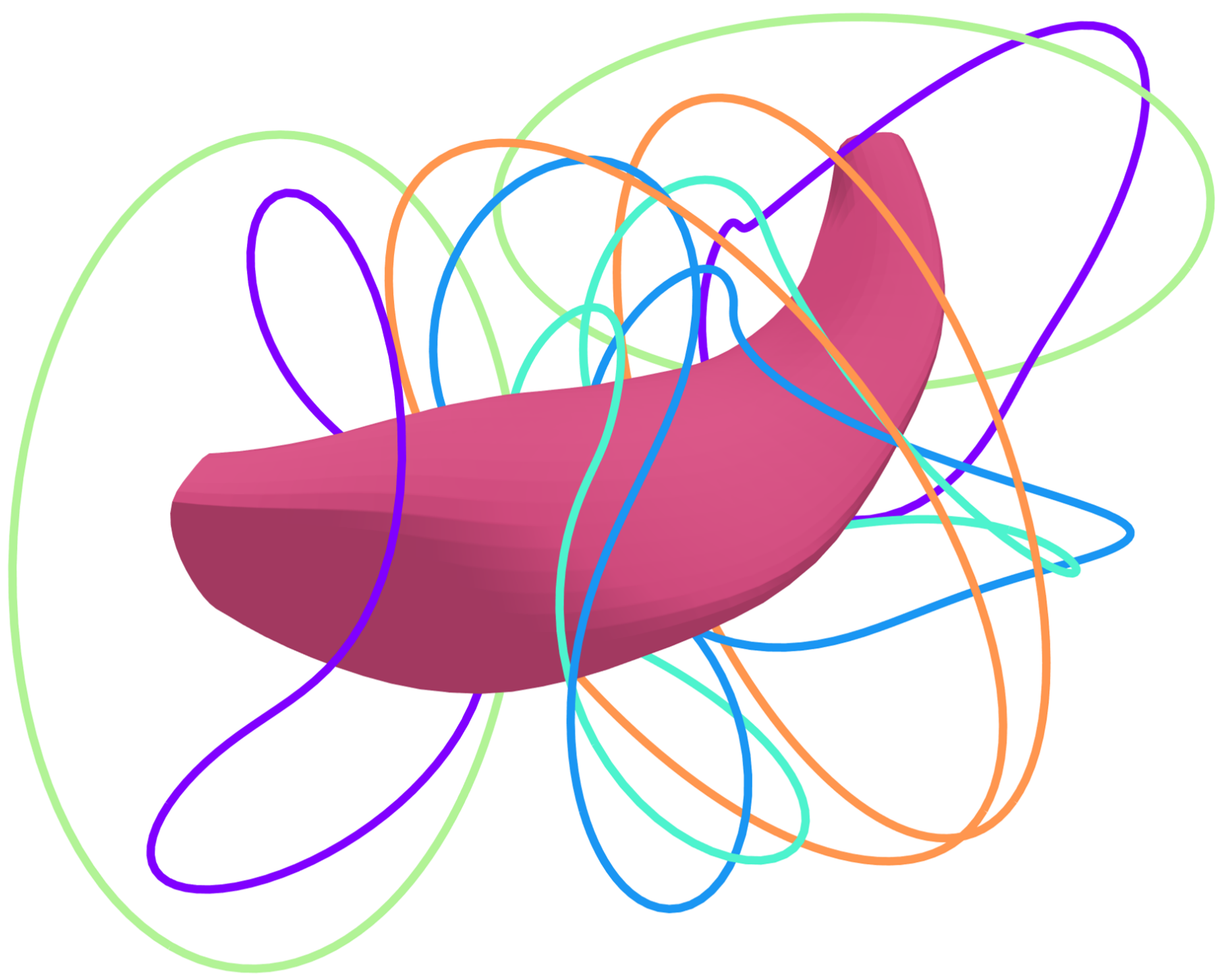}
    \caption{}
    \label{fig:frozenAllCoils}
    \end{subfigure}
    \begin{subfigure}[b]{0.25\textwidth}
    \includegraphics[width=\linewidth]{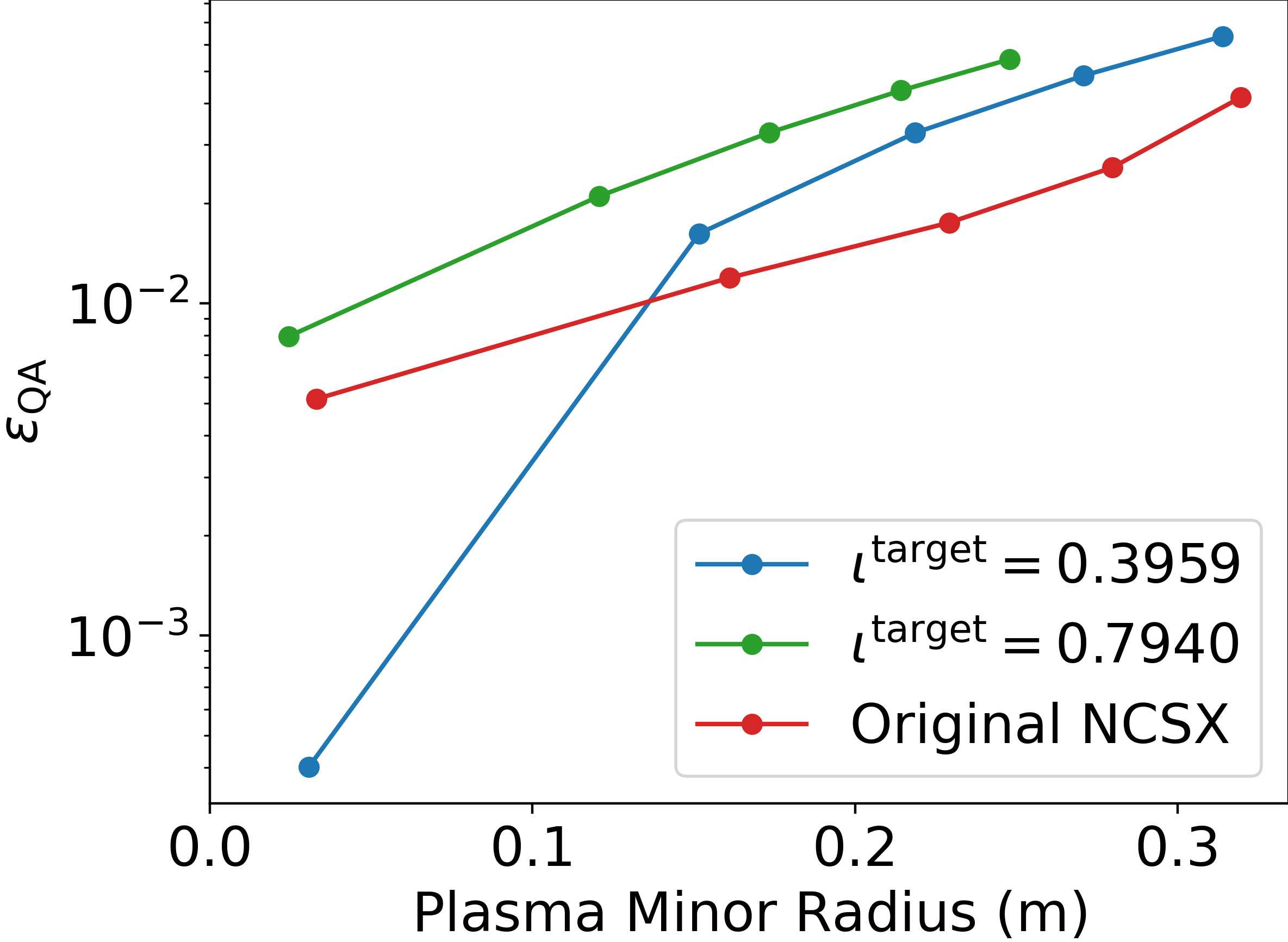}
    \caption{}
    \label{fig:frozenQA}
    \end{subfigure}
    \begin{subfigure}[b]{0.25\textwidth}
    \includegraphics[width=\linewidth]{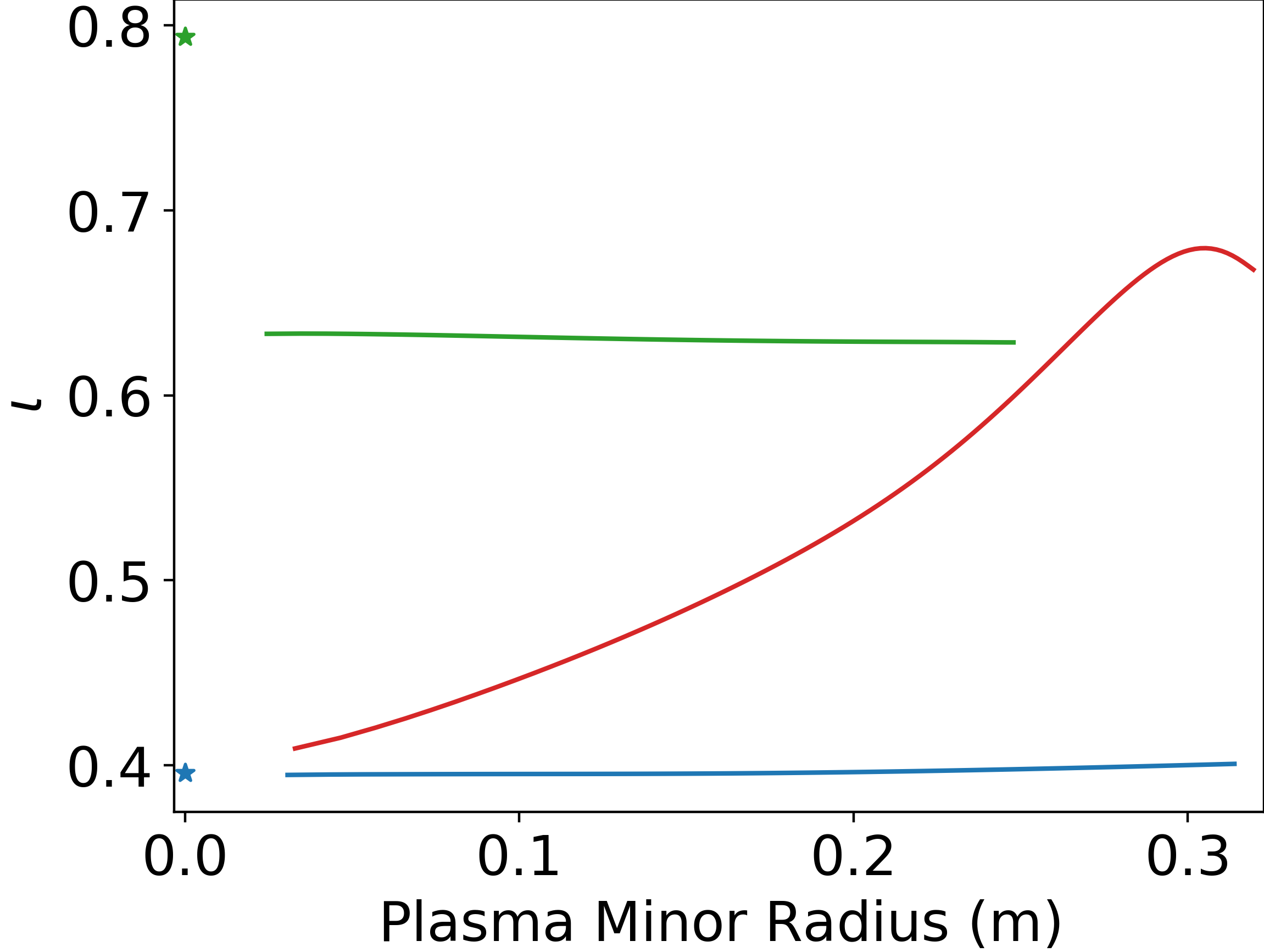}
    \caption{}
    \label{fig:frozenIota}
    \end{subfigure}
    \caption{(\subref{fig:frozenPoincare1}) and (\subref{fig:frozenPoincare2}) show Poincar\'e plots for the $\iota^{\mathrm{target}} = 0.3959$ and $\iota^{\mathrm{target}} = 0.7940$ configurations, respectively, of the device discussed in section~\ref{sec:frozen}. The (overlapping) solid blue and dotted orange curves indicate the QFM and free-boundary VMEC surfaces, respectively, used during postprocessing. The volumes of the VMEC surfaces shown are $2.958$~$\mathrm{m^3}$ (\subref{fig:frozenPoincare1}) and $1.884$~$\mathrm{m^3}$ (\subref{fig:frozenPoincare2}). 
    (\subref{fig:frozenAllCoils}) shows the modular and planar coils of the device over one field period and the VMEC boundary associated with (\subref{fig:frozenPoincare2}).
    (\subref{fig:frozenQA}) shows $\epsilon_{\mathrm{QA}}$ for each configuration of the device and the original NCSX design \cite{ncsxFlexibility}. (\subref{fig:frozenIota}) shows the rotational transform for each configuration of the device and the original NCSX design calculated using free-boundary VMEC. Stars indicate the (on-axis) optimization targets. The effective minor radius in (\subref{fig:frozenQA}) and (\subref{fig:frozenIota}) is computed with the same definition used in VMEC, given in \cite{QS3}.}
    \label{fig:frozen}
\end{figure*}

A sequential optimization consisting of two steps is useful for contextualizing the performance of the formulation outlined in the previous two sections. In the first step, modular coils alone were optimized for $\iota^{\mathrm{target}} = 0.3959$ on the axis. This was a benchmark optimization in \cite{PPO} and produces excellent quasisymmetry and nested flux surface volume. In the second step, the modular coils from the first step were frozen in place and planar coils were added as in the previous section. The currents in all the coils and the positions of the planar coils were optimized to achieve $\iota^{\mathrm{target}} = 0.7940$. A QFM surface with $\eta=50$ was added in the second step to prevent the volume of nested flux surfaces from becoming very small. As seen in figure~\ref{fig:frozen}, the nested flux surface volumes and quasisymmetry achieved using this formulation are comparable with those achieved using the formulation outlined in sections \ref{sec:formulation} and \ref{sec:magic}.
But despite only specifying two (rather than three) $\iota^{\mathrm{target}}$ values, we find less flexibility than that shown in the previous section.
This suggests that greater flexibility may be obtained by accounting for it during the initial coil optimization.

\section{Other Optimization Formulations}
\label{sec:notMagicSchemes}

In the process of coming to the result in section~\ref{sec:magic}, several optimization formulations were attempted that proved to be suboptimal. 
Without planar coils or QFM surfaces, the modular coils were rather inflexible when optimized using a sequential approach. Adding planar coils but no QFM surfaces resulted in reasonable flexibility and QA but a small plasma volume for the large $\iota$ configuration.

If multiple $\iota^{\mathrm{target}}$ were taken into account during an optimization without planar coils or QFM surfaces,
flexibility and QA could be achieved to an impressive degree, but the nested flux surface volumes were unacceptably small and the coils were too complex for practical use. Adding planar coils to this optimization created strong flexibility and QA and increased the volume of nested flux surfaces. However, the volumes were still smaller than desired, particularly for large $\iota$.
Adding QFM surfaces generated mixed results: the volume of nested flux surfaces was greatly increased, but flexibility and QA decreased. The magnitude of these effects depended strongly on the value of $\eta$.

The results surveyed in this section suggest the general, intuitive conclusion that there is a tradeoff between flexibility, quasisymmetry, and volume of nested flux surfaces. However, the optimization described in section~\ref{sec:magic} demonstrates that all three criteria can be achieved to reasonable degrees simultaneously with a single set of relatively simple coils. The existence of this coil set, which somewhat defies the tradeoff just mentioned, highlights the abundance of local minima in the objective function. Thus, future stellarator optimization studies likely should be started from many initial configurations to avoid local minima.

\section{Conclusions}

We have found that there appears to be a tradeoff between flexibility, quasisymmetry, and volume of nested flux surfaces in the context of stellarator design. 
However, we have shown an NCSX-like example device that simultaneously achieves a range of relatively large rotational transform values on the magnetic axis which are difficult to access in QA, reasonable quasisymmetry, and large nested flux surface volumes with a single, relatively simple coil set.
The most effective optimization formulation we identified involved incorporating multiple rotational transform targets, planar coil placement, and an integrability objective into the modular coil optimization.
Stellarator optimization may be improved by starting from many initial conditions to avoid local minima.

In the future, similar work could be pursued that considers windowpane coils to add primarily poloidal flux or attempts to achieve more than one type of quasisymmetry with a single coil set.
Other formulations could be examined that incorporate objectives ensuring quasisymmetry to high order off-axis \cite{QS3},
target integrability on several surfaces throughout the plasma volume,
or eliminate the need for an integrability objective by ensuring quasisymmetry on a given flux surface \cite{QSsurface}.
Stochastic optimization techniques could also be used to create coils that perform robustly even in the face of manufacturing errors \cite{stochasticOptimization}. It may be beneficial to lower the weights of the coil regularization terms. This enables coil sets to create target magnetic fields more effectively \cite{biggerCoils}, and we incidentally found that it improves flexibility. These studies could be performed for both devices that use only electromagnets and devices that use a combination of electromagnets and permanent magnets. However, substantial current changes in the modular coils were required to create flexibility in this work, which suggests that designing flexible stellarators with permanent magnets may require changing magnet arrays. This would likely only be feasible in small-scale experiments such as MUSE \cite{MUSE}.

In general, we have shown evidence that stellarators may have improved flexibility if it is accounted for in the coil design stage. It may be possible to apply similar optimization techniques to more sophisticated objective functions to optimize flexible stellarators that satisfy other important design criteria not considered here, such as MHD stability, magnetic shear, and fast ion confinement.
Packages such as SIMSOPT \cite{simsopt} will likely make attempting such optimizations relatively straightforward.
If this approach is carried out successfully, it could allow for more cost-effective experimental devices to be constructed, thus accelerating progress in the field.

\section*{Acknowledgments}
We acknowledge Florian Wechsung and Andrew Giuliani for providing the PyPlasmaOpt code. We also acknowledge conversations with Aaron Bader on the HSX control coils. We thank Neil Pomphrey for providing data from the previous NCSX flexibility study. Finally, we thank the anonymous reviewers, whose feedback was extremely helpful.

This work was made possible by funding from the Department of Energy for the Summer Undergraduate Laboratory Internship (SULI) program. This work is supported by the US DOE Contract No.{\ }DE-AC02-09CH11466. EJP was supported by the Presidential Postdoctoral Research Fellowship at Princeton University. This research was also supported by the Simons Foundation/SFARI (560651, GS and ML).

\bibliographystyle{unsrt}
\bibliography{main}

\end{document}